%% file: arxiv.tex
\documentclass[10pt,twocolumn,letterpaper]{article}

\usepackage{wacv}
\usepackage{times}
\usepackage{epsfig}
\usepackage{graphicx}
\usepackage{amsmath}
\usepackage{amssymb}
\usepackage[ruled,vlined,linesnumbered]{algorithm2e}
\usepackage{booktabs}
\usepackage{caption}
\usepackage{comment}
\usepackage{array, makecell} 
\usepackage[numbers,sort,compress]{natbib}
\usepackage{times,epsfig,graphicx,amsmath,amssymb}
\usepackage{times,graphicx,amsmath,amssymb,booktabs,tabulary,multirow,overpic,bbm,amssymb}

\wacvfinalcopy 

\ifwacvfinal
\def\assignedStartPage{1} 
\fi

\ifwacvfinal
\usepackage[breaklinks=true,bookmarks=false]{hyperref}
\else
\usepackage[pagebackref=true,breaklinks=true,colorlinks,bookmarks=false]{hyperref}
\fi

\ifwacvfinal
\else
\fi
\usepackage{authblk}
\makeatletter
\renewcommand\AB@affilsepx{, \protect\Affilfont}
\makeatother

\input{math_commands.tex}

\begin{document}

\title{A Weakly Supervised Consistency-based Learning  Method for COVID-19 Segmentation in CT Images}

\author[1,2]{Issam Laradji}
\author[2]{Pau Rodriguez}
\author[7]{Oscar Ma\~nas}
\author[3,5]{Keegan Lensink}
\author[4,5]{Marco Law}
\author[5]{Lironne Kurzman} 
\author[4,5]{William Parker}
\author[2]{David Vazquez}
\author[6]{Derek Nowrouzezahrai}

\affil[1]{issam.laradji@gmail.com}
\affil[2]{Element AI}
\affil[3]{Xtract AI}
\affil[4]{SapienML}
\affil[5]{University of British Columbia} 
\affil[6]{McGill University}
\affil[7]{Universitat Politècnica de Catalunya}

\maketitle

\begin{abstract}
Coronavirus Disease 2019 (COVID-19) has spread aggressively across the world causing an existential health crisis. Thus, having a system that automatically detects COVID-19 in tomography (CT) images can assist in quantifying the severity of the illness. Unfortunately, labelling chest CT scans requires significant domain expertise, time, and effort. We address these labelling challenges by only requiring point annotations, a single pixel for each infected region on a CT image. This labeling scheme allows annotators to label a pixel in a likely infected region, only taking 1-3 seconds, as opposed to 10-15 seconds to segment a region. Conventionally, segmentation models train on point-level annotations using the cross-entropy loss function on these labels. However, these models often suffer from low precision. Thus, we propose a consistency-based (CB) loss function that encourages  the output predictions to be consistent with spatial transformations of the input images. The experiments on 3 open-source COVID-19 datasets show that this loss function yields significant improvement over conventional point-level loss functions and almost matches the performance of models trained with full supervision with much less human effort. Code is available at: \url{https://github.com/IssamLaradji/covid19_weak_supervision}.
\end{abstract}


\section{Introduction}
The severe acute respiratory syndrome coronavirus 2 (SARS-CoV-2) has quickly become a global pandemic and resulted in over 400,469 COVID-19 related deaths as of June 8th, 2020\footnote{Source: World Health Organization.}.
The virus comes from the same family as the SARS-CoV outbreak originated in 2003 and the MERS-CoV outbreak of 2012, and is projected to join other coronavirus strains as a seasonal disease. 
The disease can present itself in a variety of ways ranging from asymptomatic to acute respiratory distress syndrome (ARDS). However, the primary and most common presentation associated with morbidity and mortality is the presence of opacities and consolidation in a patient's lungs.
As the disease spreads, healthcare centers around the world are becoming overwhelmed and facing shortages of the essential equipment necessary to manage the symptoms of the disease. 
Severe cases require admission to the intensive care unit (ICU) and need mechanical ventilation, some sources~\cite{doi:10.1056/NEJMoa2002032} citing at a rate of 5\% of all infected. Thus, availability of ICU beds due to the overwhelming number of COVID-19 cases around the world is a large challenge.
Rapid screening is necessary to diagnose the disease and slow the spread, making effective tools essential for prognostication in order to efficiently allocate intensive care services to those who need it most.

\begin{figure}[t]
  \centering
  \includegraphics[width=\linewidth,trim={0.6cm 1.2cm 0.6cm 0.6cm},clip]{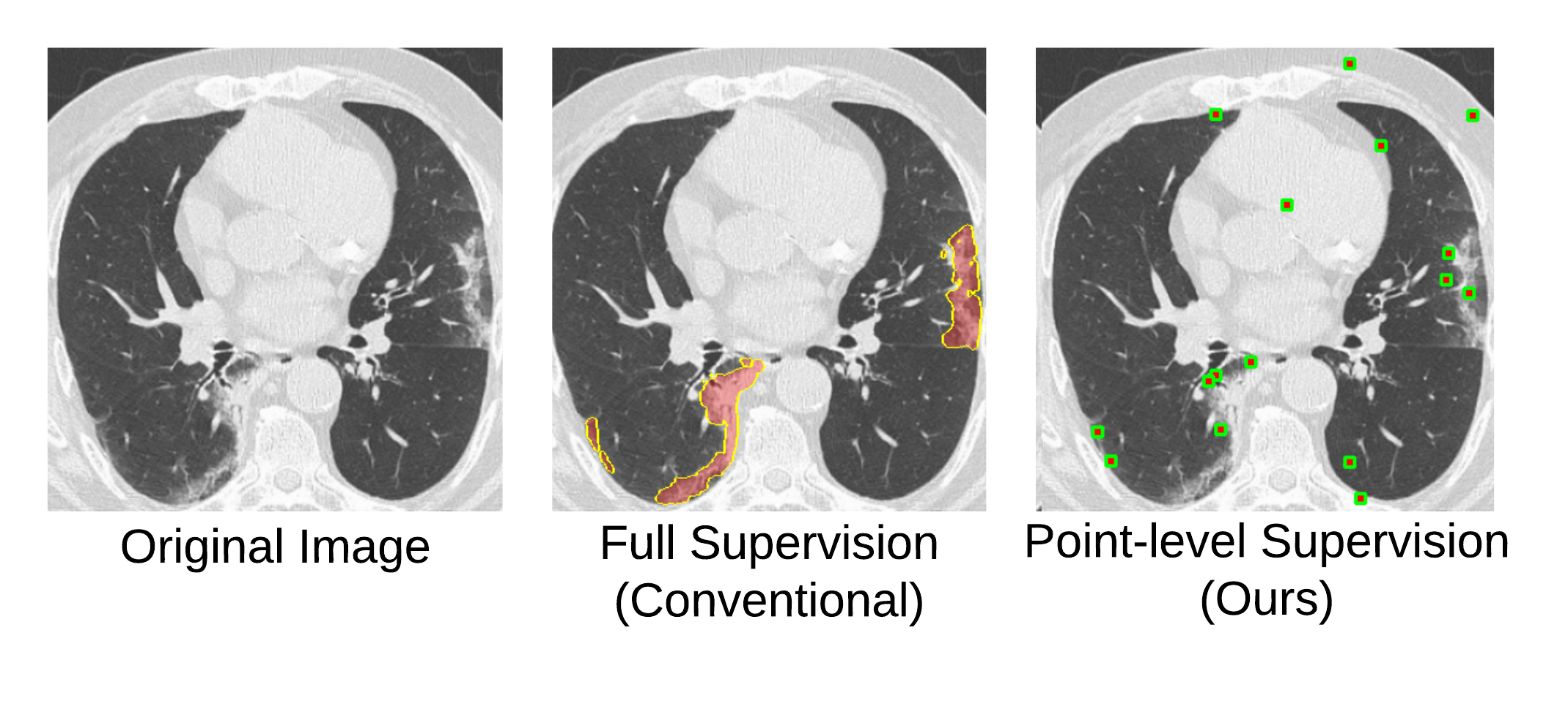}
  \caption{\textbf{Labeling Scheme.} We illustrate the difference between labels obtained using full supervision and point-level supervision. One point is placed on each infected region, and several on the background region.}
  \label{fig:model}
\end{figure}

Upon inhalation, the virus attacks and inhibits the alveoli of the lung, which are responsible for oxygen exchange~\cite{Wang2020TemporalCO}. 
In response, and as part of the inflammatory repair process, the alveoli fill with fluid, causing various forms of opacification within the lung when viewed on Computed Tomography (CT) scans. 
Due to the increased density, these areas present on CT scans as increased attenuation with preserved bronchial and vascular markings known as ground glass opacities (GGO).
In addition, the accumulation of fluid progresses to obscure bronchial and vascular regions on CT scans is known as consolidation.

While reverse transcription polymerase chain reaction (RT-PCR) has been considered the gold standard for COVID-19 screening, the shortage of equipment and strict requirements for testing environments limit the utility of this test in all settings. Further, RT-PCR is also reported to suffer from high false negative rates due to its relatively low sensitivity yet high specificity~\cite{ai2020correlation}. 
CT scans are an important complement to RT-PCR tests which were shown to demonstrate effective diagnosis, including follow-up assessment and the evaluation of disease evolution~\cite{ai2020correlation,zu2020coronavirus}.

In addition to providing complimentary diagnostic properties, the analysis of CT scans has great potential for the prognostication of patients with COVID-19. 
The percentage of well-aerated-lung (WAL) has emerged as a predictive metric for determining prognosis of patients confirmed with COVID-19, including admission to the ICU and death~\cite{Colombi2020WellaeratedLO}.
The quantification of percentage of WAL is often done by visually estimating volume of opacification relative to healthy lung, and can be estimated automatically through attenuation values within the lung.
In addition to the percent of WAL, which does not account for the various forms of opacification, expert interpretation of CT scans can provide insight on the severity of the infection by identifying various patterns of opacification. 
The prevalence of these patterns, which are correlated with the severity of the infection, has been correlated to different stages of the disease~\cite{Li2020,Wang2020}. 
The quantification of both the percentage of WAL and the opacification composition enables efficient estimation of the stage of the disease and the patient outcome.


Deep learning-based methods have been widely applied in medical image analysis to combat COVID-19~\cite{keshani2013lung, hesamian2019deep, wang2017central}. They have been proposed to detect patients infected with COVID-19 via radiological imaging. For example,  COVID-Net~\cite{wang2020covid} was proposed to detect COVID-19 cases from chest radiography images. An anomaly detection model was designed to assist radiologists in analyzing the vast amounts of chest X-ray images~\cite{schlegl2017unsupervised}. For CT imaging, a location-attention oriented model was employed to calculate the infection probability of COVID-19~\cite{butt2020deep}. A weakly-supervised deep learning-based software system was developed in~\cite{zheng2020deep} using 3D CT volumes to detect COVID-19. A list of papers for COVID-19 imaging-based AI works can be found in \citet{wang2020cord}. Although plenty of AI systems have been proposed to provide assistance in diagnosing COVID-19 in clinical practice, there are only a few related works~\cite{fan2020inf}, and no significant impact has been shown using AI to improve clinical outcomes, as of yet.

According to \citet{ma2020towards}, it takes around 400 minutes to delineate one CT scan with 250 slices. That is an average of 1.6 minutes per slice. On the other hand, it takes around 3 seconds to point to a single region at the pixel level~\citet{papadopoulos2017training}. Thus, point-level annotations allow us to label many more slices quickly.

Point-level annotations are not as expressive as segmentation labels, making effective learning a challenge for segmentation models (Fig.~\ref{fig:model}). Conventionally, segmentation models train on point-level annotations using the cross-entropy on these labels. While this loss can yield good results in some real-life datasets~\cite{bearman2016s}, the resulting models usually suffer from low precision as they often predict big blobs. Such predictions are not suitable for imbalanced images where only few small regions are labeled as foreground. Thus, we propose a consistency-based (CB) loss function that encourages the model's output predictions to be consistent with spatial transformations of the input images. While consistency methods have been successfully deployed in semantic segmentation, the novel aspect of this work is the notion of consistency under weak supervision, which utilizes unlabeled pixels during training. We show that this regularization method yields significant improvement over conventional point-level loss functions was on 3 open-source COVID-19 datasets. We also show that this loss function results in a segmentation performance that almost matches that of the fully supervised model. To the best of our knowledge, this is the first time that self-supervision has been applied in conjunction with point-level supervision on a medical segmentation dataset.

We summarize our contributions and results on 3 publicly available CT Scans~\footnote{Found here: https://medicalsegmentation.com/covid19/} as follows:
\begin{enumerate}
    \setlength\itemsep{0mm}
    \item We propose a framework that trains using a consistency-based loss function on a medical segmentation dataset labeled with point-level supervision.
    \item We present a trivial, yet cost-efficient point-level supervision setup where the annotator is only required to label a single point on each infected region and several points on the background.
    \item We show that our consistency-based loss function yields significant improvement over conventional point-level loss functions and almost matches the performance of models trained with full supervision.
\end{enumerate}

\section{Related Work}
\label{sec:related_work}
In this section, we start with reviewing semantic segmentation methods applied to CT scans on general medical problems, followed by semantic segmentation for COVID-19. Later we go over semantic segmentation methods for weakly supervised problem setups and self-supervision methods that were shown to help generalization performance for semantic segmentation.

\paragraph{Semantic segmentation for CT Scans} has been widely used for diagnosing lung diseases. Diagnosis is often based on segmenting different organs and lesions from chest CT slices, which can provide essential information for doctors to identify lung diseases. Many methods exist that perform nodule segmentation of lungs. Early algorithms are based on image processing and SVMs to segment nodules~\cite{keshani2013lung}. Then, algorithms based on deep learning emerged~\cite{hesamian2019deep}. These methods include central focus CNNs~\citep{wang2017central} and GAN-synthesized data for nodule segmentation in CT scans~\cite{jin2018ct}. A recent method uses multiple deep networks to segment lung tumors from CT slices with varying resolutions, and multi-task learning of joint classification and segmentation~\cite{jiang2018multiple}. In this work, we use an ImageNet-pretrained FCN8~\cite{long2015fully} as our segmentation method.

\paragraph{Semantic segmentation for COVID-19} 
While COVID-19 is a recent phenomenon, several methods have been proposed to analyze infected regions of COVID-19 in lungs. \citet{fan2020inf} proposed a semi-supervised learning algorithm for automatic COVID-19 lung infection segmentation from CT scans. Their algorithm leverages attention to enhance representations. Similarly, \citet{zhou2020automatic} proposed to use spatial and channel attention to enhance representations, and \citet{chen2020residual} augment U-Net~\cite{Ronneberger2015} with ResNeXt~\citep{xie2017aggregated} blocks and attention. Instead of focusing on the architecture, \citet{amyar2020multi} proposed to improve the segmentation performance with a multi-task learning approach which includes a reconstruction loss. Although previous methods are accurate, their computational cost can be prohibitive. Thus, \citet{qiu2020miniseg} proposed Miniseg for efficient COVID-19 segmentation. Unfortunately, these methods require full supervision, which is costly to acquire compared to point-level supervision: our problem setup.

\paragraph{Weakly supervised semantic segmentation} methods can vastly reduce the required annotation cost for collecting a training set. According to \citet{bearman2016s}, manually collecting image-level and point-level labels for the PASCAL VOC dataset~\citep{everingham2010pascal} takes only $20.0$ and $22.1$ seconds per image, respectively. These annotation methods are an order of magnitude faster than acquiring full segmentation labels, which is $239.0$ seconds on average. Other forms of weaker labels were explored as well, including bounding boxes~\cite{khoreva2017simple} and image-level annotation~\cite{Zhou2018PRM}. Weak supervision was also explored in instance segmentation where the goal is to identify object instances as well as their class labels~\cite{laradji2019instance,laradji2019masks,zhou2018weakly}. In this work, the labels are given as point-level annotations instead of the conventional per-pixel level labels and the task is to identify the class labels of the regions only.

\paragraph{Self-supervision for weakly supervised} semantic segmentation is a relatively new research area that has strong potential in improving segmentation performance. The basic idea is to generate two perturbed versions of the input and apply consistency training to encourage the predictions to be similar~\cite{xie2019unsupervised}. For example, FixMatch~\cite{sohn2020fixmatch} combined consistency regularization with pseudo-labeling to produce artificial image-level labels. In the case of dense predictions, the outputs need to be further transformed in order to compare them against a consistency loss, making the model's output equivariant against transformations. Self-supervision was recently applied in a weakly supervised setup where annotations are image-level~\cite{wang2020selfsupervised}.  The idea was to make the output consistent across scales, which led to new state-of-the-art results on PASCAL VOC dataset. \citet{ouali2020semisupervised} proposed to apply cross-consistency training, where the perturbations are applied to the outputs of the encoder and the dense predictions are enforced to be invariant. These perturbations can also be used for data augmentation, which can be learnt automatically using methods based on reinforcement learning and bilevel optimization~\cite{cubuk2018autoaugment,mounsaveng2020learning}. For medical segmentation, self-supervision has been used along with semi-supervised learning~\cite{bortsova2019semi, Li_2020}. \citet{bortsova2019semi} made the outputs consistent across elastic transforms, while \citet{Li_2020} added a teacher-student paradigm for consistency training. In this work, we apply consistency loss on the novel setup of medical segmentation with point supervision.

\section{Methodology}
\label{sec:methodology}

\begin{figure*}[!t]
\centering
\includegraphics[width=1.0\textwidth,trim={0 0.4cm 0 0.4cm},clip]{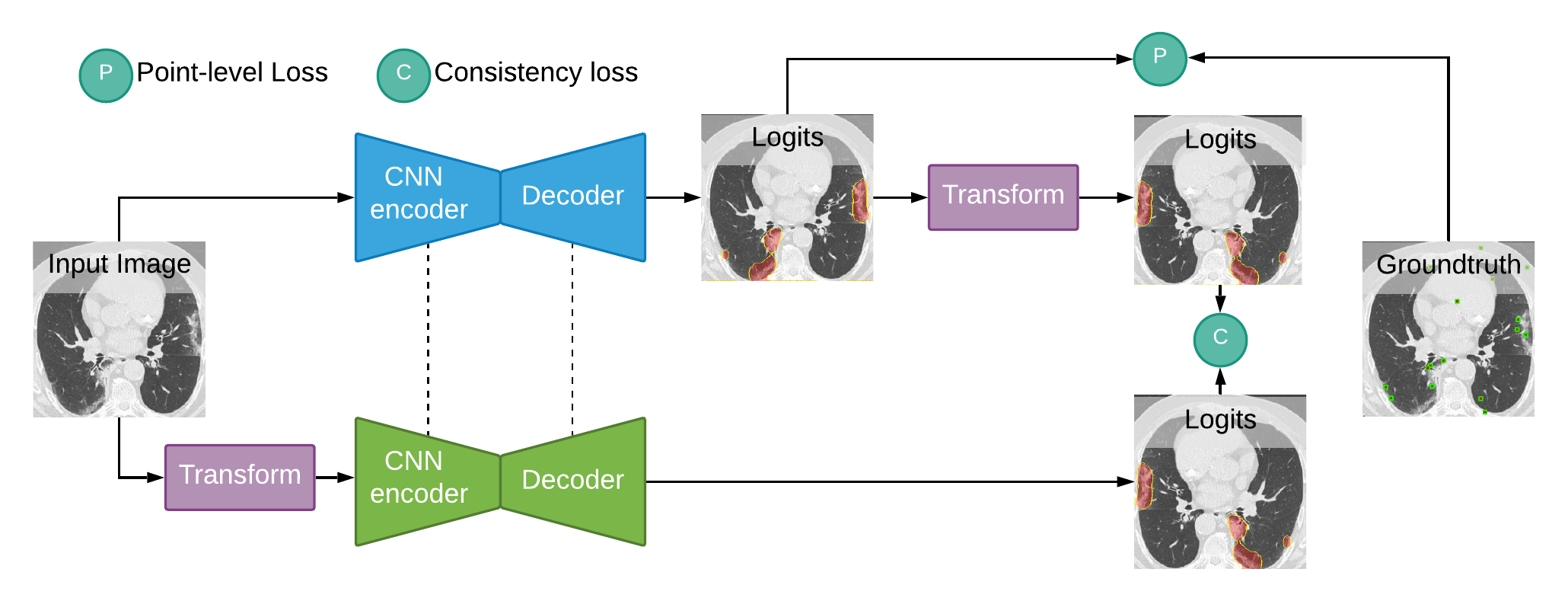}
\caption{{\bf Model Training.} Our model has two branches with shared weights. The first branch encodes the original input $x$ while the second branch encodes the transformed input $t(x)$. The point-level loss compares the outputs $f(x)$ and $f(t(x))$ with the corresponding weak labels $y$ and $t(y)$. In addition, an unsupervised consistency loss is used to make the outputs $t(f(x))$ and $f(t(x))$ consistent.}
\label{fig:model2}
\end{figure*}

\paragraph{Problem Setup and Network Architecture.} We define the problem setup as follows. Let $X$ be a set of $N$ training images with corresponding ground truth labels $Y$. $Y_i$ is a $W \times H$ matrix with non-zero entries that indicate the locations of the object instances. The values of these entries indicate the class label that the point corresponds to.

We use a standard fully-convolutional neural network that takes as input an image of size $W \times H$ and outputs a $W \times H \times C$ per-pixel map where $C$ is the set of object classes of interest. The output map is converted to a per-pixel probability matrix $S_i$ by applying the softmax function across classes. These probabilities indicate how likely each pixel belongs to the infected region of a class $c \in C$.

\paragraph{Proposed Loss Function.} 
Our weakly supervised method uses a loss function that consists of a supervised point-level loss and an unsupervised consistency loss. Given a network $f_\theta$ that outputs a probability map $S_i$ given an image $X_i$, we optimize its parameters $\theta$ using the following loss function,

\begin{equation}
\mathcal{L}(X, Y) = \sum_{i=1}^N \underbrace{\mathcal{L}_P(X_i,Y_i)}_{\text{Point-level}} + \lambda\underbrace{\mathcal{L}_C(X_i)}_{\text{Consistency}}\;,
\label{eq:loss}
\end{equation}
where $\lambda$ is used to weigh between the two loss terms.

\paragraph{Point-level loss.} We apply the standard cross-entropy function against point annotations, which is defined as follows,
\begin{equation} \label{eq:pointlevel}
\mathcal{L}_P(X_i,Y_i) = -\sum_{j \in \mathcal{I}_i}\log(f_\theta(X_i)_{jY_j})\;,
\end{equation}
where $f_\theta(X_i)_{jY_j}$ is the output corresponding to class $Y_j$ for pixel $j$, and $\mathcal{I}_i$ is the set of labeled pixels for image $X_i$.

\paragraph{Consistency loss.} We first define a set of geometric transformations $T=\{t_1, t_2, ..., t_n\}$. An example of $t_k$ is horizontal flipping, which can be used to transform an image $X_i$ and its corresponding label $Y_i$ collectively to their flipped version. The goal of this loss function  is to make the model's output consistent with respect to these transformations on the input image. The loss function is defined as follows,
\begin{equation}\label{eq:cons}
    \mathcal{L}_C(X_i) = \sum_{j\in \mathcal{P}_i} | t_k(f_\theta(X_i))_{j} - f_\theta(t_k(X_i))_{j} |,
\end{equation}
where $\mathcal{P}_i$ is the set of pixels for image $X_i$. This unsupervised loss function helps the network learn equivariant semantic representations that go beyond the  translation equivariance that underlies convolutional neural networks, serving as an additional form of supervision. 

\paragraph{Model Training.}
The overview of the model training is shown in Fig.~\ref{fig:model2} and Alg.~\ref{algo1}. The model has two branches with shared weights $\theta$. At each training step $k$, we sample an image $X_i$ and a transform function $t_k \in T$. The model's first branch takes as input the original image $X_i$ and the second branch takes as input the transformed image $t_k(X_i)$. The transformed output of the first branch, $y_1 := t_k(f_\theta(X_i))$, is aligned with the prediction of the second branch $y_2 := f_\theta(t_k(X_i))$ for pixel-wise comparison by the consistency loss function~\ref{eq:cons}.

In addition to the consistency loss, the point-level loss $\mathcal{L}_P$ is applied to both input $X_i$ and $t_k(X_i)$, \ie $\mathcal{L}_P(t_k(X_i),t_k(Y_i))$, where $t_k(Y_i)$ is a pseudo ground-truth mask for $t_k(X_i)$ generated by applying the same geometric transformation $t_k$ to the ground-truth mask $Y_i$. In this case, the network is forced to update the prediction for $t_k(X_i)$ to be more similar to $t_k(Y_i)$.

In this work, we use geometric transformations which allow us to infer the true label of images that undergo these transformations. For instance, the segmentation mask of the flipped version of an image is the flipped version of the original segmentation mask. Thus, we include the following transformations: 0, 90, 180 and 270 degree rotation and a horizontal flip. At test time, the trained model can then be directly used to segment infected regions on unseen images with no additional human input.

\newcommand\mycommfont[1]{\footnotesize\ttfamily{#1}}
\SetCommentSty{mycommfont}

\begin{algorithm}
\caption{Model Training}
\label{algo1}
\DontPrintSemicolon
\SetAlgoLined
\SetKwInOut{Input}{Input}
\SetKwInOut{Output}{Output}
\SetKwInOut{Parameter}{Parameters}
\Input{$X = \{X_1, X_2, ..., X_n\}$ images, $Y = \{Y_1, Y_2, ..., Y_n\}$
point-level masks.}
\Output{Trained parameters $\theta^*$}
\Parameter{A weight coefficient $\lambda$,\\
A set of transformation functions $T$,\\
A model forward function $f_\theta$.}
\BlankLine

\For{each batch $B$} {
$\mathcal{L} \gets 0$\\
\For{each $(X_i, Y_i) \in B$} {
\emph{Compute Point Loss}\\
$\mathcal{L}_P \gets -\sum_{j \in \mathcal{I}_i}\log(f_\theta(X_i)_{jY_j})$\\[0.1in]
\emph{Uniformly sample a transform function}\\
$t_k \sim T$\\[0.1in]
\emph{Compute Consistency Loss}\\
$\mathcal{L}_C \gets \sum_{j\in \mathcal{P}_i} | t_k(f_\theta(X_i))_{j} - f_\theta(t_k(X_i))_{j} |$\\[0.1in]
$\mathcal{L} \gets \mathcal{L} + \mathcal{L}_P + \lambda \mathcal{L}_C$\\
}
Update $\theta$  by backpropagating w.r.t. $\mathcal{L}$
}
\end{algorithm}

\section{Experiments}
\label{sec:experiments}

\begin{table*}
\caption{Statistics of open-source COVID-19 datasets.}
\label{tab:datasets}
\centering
\begin{tabular}{lrrrr}
\toprule
       Name &  \# Cases &  \# Slices  &  \# Slices with Infections (\%) &  \# Infected Regions \\
\midrule
 COVID-19-A &    60 &       98 &                        98 (100.0\%) &                 776 \\
 COVID-19-B &    9 &        829 &                       372 (44.9\%) &                1488 \\
 COVID-19-C &    20 &       3520 &                     1841 (52.3\%) &                5608 \\
\bottomrule
\end{tabular}
\end{table*}

\subsection{Experimental Setup}
Here we describe the details behind the datasets, methods, and evaluation metrics used in our experiments.

\subsubsection{Datasets}
\label{ssec:setup}
We evaluate our weakly supervised learning system on three separate open source medical segmentation datasets (referred to as COVID-19-A/B/C). For each dataset, a point-level label is obtained for a segmentation mask by taking the pixel with the largest distance transform as the centroid. 

Thus, we generate a single supervised point for each disjoint infected region on the training images. For the background region, we randomly sample several pixels as the ground-truth points (Figure~\ref{fig:model}). We show  the dataset statistics in Table~\ref{tab:datasets} and describe them in the next sections.

\paragraph{COVID-19-A~\cite{covid19a, fan2020inf}} consists of 100 axial lung CT JPEG images obtained from 60 COVID-19 lung CTs provided by the Italian Society of Medical and Interventional Radiology. Each image was labeled for ground-glass, consolidation, and pleural effusion by a radiologist. We discarded two images without areas of infection from this dataset due to their low resolution. Images were resized to a fixed dimension of $352 \times 352$ pixels and normalized using ImageNet statistics~\cite{ILSVRC15}. The final dataset consisted of 98 images separated into a training set ($n=50$), validation set ($n=5$), and a test set ($n=48$).

\paragraph{COVID-19-B~\cite{covid19a}} consists of 9 volumetric COVID-19 chest CTs in DICOM format containing a total of 829 axial slices. Images were first converted from Houndsfield units to unsigned 8-bit integers, then resized to $352 \times 352$ pixels and normalized using ImageNet statistics~\cite{ILSVRC15}. 

We use COVID-19-B to evaluate the consistency loss on two splits of the dataset: separate and mixed. In the separate split (COVID-19-B-Separate), the slices in the training, validation, and test set come from different scans. The goal is to have a trained model that can generalize to scans of new patients. In this setup, the first 5 scans are defined as the training set, the sixth scan as validation, and the remaining scans as the test set. 

For the mixed split (COVID-19-B-Mixed), the slices in the training, validation, and test set come from the same scans. The idea is to have a trained model that can infer the masks in the remaining slices of a scan when the annotator only labels few of the slices in that scan.  In this setup, the first 5 scans are defined as the training set, the sixth scan as validation, and the remaining scans as the test set. For each scan, the first 45\% slices of the scan are defined as the training set, the next 5\% as the validation set, and the remaining slices as the test set.

\paragraph{COVID-19-C~\cite{zenobo}} consists of 20 CT volumes. Lungs and areas of infection were labeled by two radiologists and verified by an experienced radiologist. Each three-dimensional CT volume was converted from Houndsfield units to unsigned 8-bit integers and normalized using ImageNet statistics~\cite{ILSVRC15}. 

As with COVID-19-B, we also split the dataset into {\it separate} and {\it mixed} versions to evaluate our model's efficacy. For the separate split (COVID-19-B-Sep), we assign 15 scans to the training set, 1 scan to the validation set, and 4 scans to the test set. For the mixed split (COVID-19-C-Mixed), we separate the slices from each scan in the same manner as in COVID-19-B, training on the first $45\%$ axial slices, validating on the next $5\%$ of slices, and testing on the remaining $50\%$ of slices.

\subsubsection{Evaluation Metrics}
As common practice~\cite{shan2020lung}, we evaluate our models against the following metrics for semantic segmentation:
\vspace{-3mm}
\paragraph{Intersection over Union (IoU)} measures the overlap between the prediction and the ground truth: 
$IoU = \frac{TP}{TP + FP + FN},$
where TP, FP, and FN is the number of true positive, false positive and false negative pixels across all images in the test set.
\vspace{-3mm}
\paragraph{Dice Coefficient (F1 Score)} is similar to IoU but gives more weight to the intersection between the prediction and the ground truth: $F1 = \frac{2 * TP}{2 * TP + FP + FN}.$
\vspace{-3mm}
\paragraph{PPV (Positive Predicted Value)} measures the fraction of positive samples that were correctly predicted, which is also known as precision: $PPV = \frac{TP}{TP + FP}$.
\vspace{-3mm}
\paragraph{Sensitivity (recall)} measures the fraction of real positive samples that were predicted correctly: $Sensitivity = \frac{TP}{TP + FN}$.
\vspace{-3mm}
\paragraph{Specificity (true negative rate)} measures the fraction of real negative samples that were predicted correctly: $Specificity = \frac{TN}{FP + TN}$.

\subsection{Methods and baselines}
We provide experiments with three weakly supervised loss functions based on point-level annotations and a fully-supervised upper bound method:

\begin{itemize}
    \setlength\itemsep{0mm}
    \item \textit{Point loss (PL)}. It is defined in Eq. 2 in \citet{bearman2016s}. The loss function encourages all pixel predictions to be background for background images and applies cross-entropy against the provided point-level annotations, ignoring the rest of the pixels.
    \item \textit{CB(Flip) + PL}. It is defined in Eq.~\ref{eq:loss} in Section~\ref{sec:methodology}, which combines the point loss and the horizontal flip transformation for the consistency loss.
    \item \textit{CB(Flip, Rot) + PL}. It is the same as \textit{CB(Flip) + PL} except that the transformation used for the consistency loss also includes the 0, 90, 180, and 270 degree rotation transformation uniformly sampled for each image.
    \item \textit{Fully supervised}. This loss function combines weighted cross-entropy and IoU loss as defined in Eq. (3) and (5) from~\citet{wei2019f3net}, respectively. It is an efficient method for ground truth segmentation masks that are imbalanced. Since this loss function requires full supervision, it serves as an upper bound performance in our experimental results.
\end{itemize}

\paragraph{Implementation Details}
Our methods use an Imagenet-pretrained VGG16 FCN8 network~\cite{long2015fully}. Models are trained with a batch size of 8 for 100 epochs with ADAM~\cite{kingma2014adam} and a learning rate of $10^{-4}$. We also achieved similar results with optimizers that do not require a learning rate~\cite{vaswani2019painless,loizou2020stochastic,vaswani2020adaptive}. The reported scores are on the test set which were obtained with early stopping on the validation set. Point annotations were obtained by uniformly sampling one pixel from each annotated mask. The same amount of points are uniformly sampled from the background.

\begin{figure*}[tbh]
  \centering
  \includegraphics[width=0.93\linewidth]{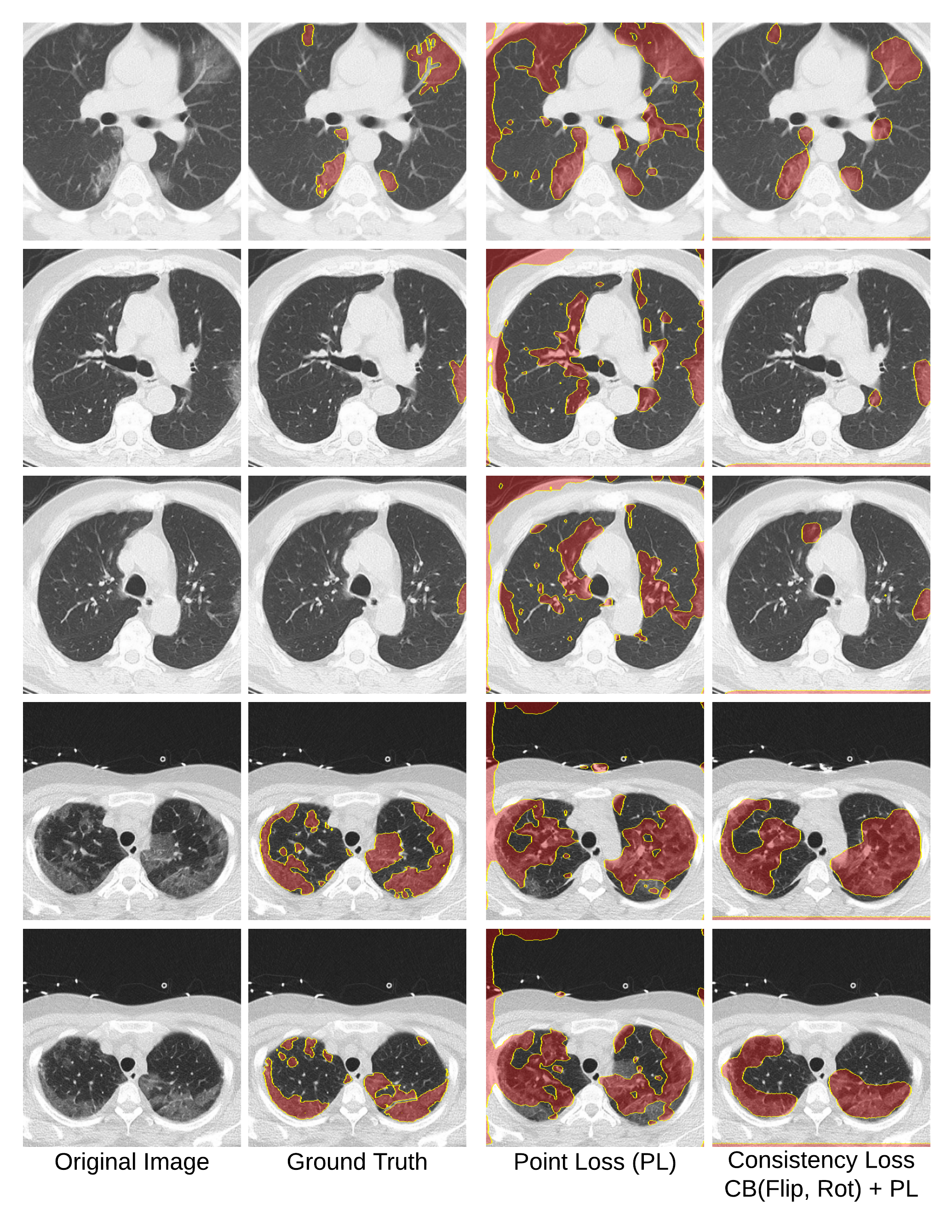}
  \caption{\textbf{Qualitative results.} We show the predictions obtained from training the model with the point-level loss in~\citet{bearman2016s} and our consistency-based (CB) loss. With the CB loss the predictions are much closer to the ground-truth labels.}
  \label{fig:qualitative}
\end{figure*}

\subsection{Segmentation Results}
Here we evaluate the loss functions on three covid datasets and discuss their results. 

\subsubsection{COVID-19-A}
\label{sssec:covid-19-a}
Table~\ref{tab:covid19a} shows that with only point supervision, our method was able to perform competitively compared to full supervision. In terms of sensitivity, it can be observed that the point loss outperformed the fully-supervised baseline by 0.11 points. For the other metrics, we were able to obtain competitive performance when using the consistency-based (CB) loss.  The gap between fully supervised and point-based loss is reduced when using flips and rotations (Flip, Rot) instead of simple horizontal flips (Flip). Moreover, with (Flip, Rot), our method surpasses the fully-supervised sensitivity by 0.12 points. COVID-19-A is a small and easy dataset compared to COVID-19-B and COVID-19-C. Thus, in the next sections, we show that with bigger datasets, \textit{CB point loss} obtains even better performance on the rest of the metrics with weak supervision.

\begin{table}
\centering
\caption{COVID-19-A Segmentation Results}
\label{tab:covid19a}
\resizebox{\columnwidth}{!}{%
\begin{tabular}{lrrrrr}
\toprule
    Loss Function &  Dice &  IoU &  PPV &  Sens. &  Spec. \\
\midrule
    Fully Supervised &  0.65 & 0.48 & 0.52 &   0.85 &   0.85 \\
    Point Loss (PL) & 0.54 & 0.37 & 0.39 & \textbf{0.88} &   0.73 \\
    CB(Flip) + PL (Ours) &  0.58 & 0.41 & 0.46 &   0.80 &   0.82 \\
    CB(Flip, Rot) + PL (Ours) &  \textbf{0.73} & \textbf{0.57} & \textbf{0.65} &   0.82 & \textbf{0.92} \\
\bottomrule
\end{tabular}
}
\end{table}

\subsubsection{COVID-19-B}
\label{sssec:covid-19-b}
As seen in Table~\ref{tab:covid19b-mixed} and \ref{tab:covid19b-sep}, the CB method is more robust against different splits of the data. In both COVID-19-B-Sep and COVID-19-B-Mixed, the CB method achieves similar results, whereas there is more variance in the results with \textit{Point Loss} and \textit{W-CE} metrics. While the W-CE baseline has an average gap of 0.37 between \textit{sep} and \textit{mixed} over all metrics, the CB Point loss only has a difference of 0.07 with (Flip) and 0.08 with (Flip, Rot). Remarkably, on \textit{sep}, our weakly supervised method with (Rot, Flip) improved by 0.48,  0.42, and 0.56, on the Dice, IoU, and Sensitivity metrics, with respect to the W-CE baseline. On PPV and Specificity, our method was able to retain a competitive performance, with a difference of 0.16 and 0.02 respectively. Except the for Sensitivity in COVID-19-B-Sep, the CB loss (Rot, Flip) yields better results than the point loss.

\begin{table}
\centering
\caption{COVID-19-B-Mixed Segmentation Results}
\label{tab:covid19b-mixed}
\resizebox{\columnwidth}{!}{%
\begin{tabular}{lrrrrr}
\toprule
    Loss Function &  Dice &  IoU &  PPV &  Sens. &  Spec. \\
\midrule
    Fully Supervised &  0.84 & 0.73 & 0.90 &   0.80 &   1.00 \\
    Point Loss (PL) &  0.33 & 0.20 & 0.20 &   0.91 &   0.94 \\
    CB(Flip) + PL (Ours) &  0.73 & 0.57 & \textbf{0.64} &   0.85 &   \textbf{0.99} \\
    CB(Flip, Rot) + PL (Ours) &  \textbf{0.75} & \textbf{0.60} & 0.63 &   \textbf{0.92} &   \textbf{0.99} \\
\bottomrule
\end{tabular}
}
\end{table}

\begin{table}
\centering
\caption{COVID-19-B-Sep Segmentation Results}
\label{tab:covid19b-sep}
\resizebox{\columnwidth}{!}{%
\begin{tabular}{lrrrrr}
\toprule
    Loss Function &  Dice &  IoU &  PPV &  Sens. &  Spec. \\
\midrule
    Fully Supervised &  0.24 & 0.14 & 0.89 &   0.14 &   1.00 \\
    Point Loss (PL) &  0.57 & 0.40 & 0.44 &   \textbf{0.82} &   0.94 \\
    CB(Flip) + PL (Ours) &  0.69 & 0.53 & 0.72 &   0.66 &   \textbf{0.99} \\
    CB(Flip, Rot) + PL (Ours) &  \textbf{0.72} & \textbf{0.56} & \textbf{0.73} &   0.70 &   0.98 \\
\bottomrule
\end{tabular}
}
\end{table}

\subsubsection{COVID-19-C}
As seen in Tables \ref{tab:covid19c-mixed} and \ref{tab:covid19c-sep}, the fully supervised method performs better on COVID-19-C than in the other two datasets and the performance gap between mixed and sep is smaller. This can be attributed to the larger size of COVID-19-C. The average gap in performance of the fully supervised baseline between the \textit{mixed} and \textit{sep} versions is 0.06 for COVID-19-C. The weakly supervised CB loss yields a gap of 0.05 in performance between \textit{mixed} and \textit{sep}. Similar to COVID-19-B, except for Sensitivity, the CB point loss yields substantially better results than the point loss. We also observed better results when adding rotations. In fact, with (Flip, Rot), our weakly supervised method improves over the fully supervised baseline by 0.04, 0.04, and 0.21 on Dice, IoU and \textit{Sensitivity} on the \textit{sep} split.

\begin{table}
\centering
\caption{COVID-19-C-Mixed Segmentation Results}
\label{tab:covid19c-mixed}
\resizebox{\columnwidth}{!}{%
\begin{tabular}{lrrrrr}
\toprule
    Loss Function &  Dice &  IoU &  PPV &  Sens. &  Spec. \\
\midrule
    Fully Supervised &  0.78 & 0.64 & 0.79 &   0.77 &   1.00 \\
    Point Loss (PL) &  0.12 & 0.07 & 0.07 &   \textbf{0.95} &   0.82 \\
    CB(Flip) + PL (Ours) &  0.66 & 0.49 & \textbf{0.56} &   0.80 &   \textbf{0.99} \\
    CB(Flip, Rot) + PL (Ours) &  \textbf{0.68} & \textbf{0.51} & \textbf{0.56} &   0.85 &   \textbf{0.99} \\
\bottomrule
\end{tabular}
}
\end{table}

\begin{table}
\centering
\caption{COVID-19-C-Sep Segmentation Results}
\label{tab:covid19c-sep}
\resizebox{\columnwidth}{!}{%
\begin{tabular}{lrrrrr}
\toprule
    Loss Function &  Dice &  IoU &  PPV &  Sens. &  Spec. \\
\midrule
    Fully Supervised &  0.71 & 0.55 & 0.78 &   0.65 &   0.99 \\
    Point Loss (PL) &  0.37 & 0.23 & 0.23 &   \textbf{0.97} &   0.76 \\
    CB(Flip) + PL (Ours) &  0.69 & 0.53 & 0.62 &   0.79 &   0.96 \\
    CB(Flip, Rot) + PL (Ours) &  \textbf{0.75} & \textbf{0.59} & \textbf{0.66} &   0.86 &   \textbf{0.97} \\
\bottomrule
\end{tabular}
}
\end{table}

\subsection{Counting and Localization Results}
In this setup we consider the task of counting and localizing COVID-19 infected regions in CT Scan images. Radiologists strive to identify all regions that might have relevance to COVID-19, which is a very challenging task, especially for small infected regions. Thus, having a model that can localize these regions can help improve radiologist performance in the identification of infected regions.

We consider the COVID-19-B and COVID-19-C datasets to evaluate 3 types of loss functions: point loss (Eq.2 from~\citet{bearman2016s}), LCFCN loss (Eq. 1 from~\citet{laradji2018blobs}), and consistency-based LCFCN loss that we propose in this section. 

The consistency based LCFCN (CB LCFN Loss) loss extends the LCFCN loss with the CB loss proposed in Eq.~\ref{eq:loss} using the horizontal flip transformation. To evaluate these 3 loss functions, we consider each connected infected region as a unique region. The goal is to identify whether these regions can be counted and localized. We use the mean absolute error (MAE) and grid average mean absolute error (GAME)~\citep{guerrero2015Trancos} to measure how well the methods can count and localize infected regions. We provide results for $GAME(L=4)$ which divides the image using a grid of $4^L$ non-overlapping regions, and the error is computed as the sum of the MAE in each of these subregions.

Table~\ref{tab:loc1} and \ref{tab:loc2} shows that the consistency loss helps LCFCN achieve superior results in counting and localizing infected regions in the CT image. It is expected that the \textit{Point Loss} achieves poor performance as it predicts big blobs that can encapsulate several regions together. On the other hand, the consistency loss helped LCFCN improve its results suggesting the model learns more informative semantic features for the task with such self-supervision.

\begin{table}
\centering
\caption{COVID-19-B-Mixed Counting and Localization}
\label{tab:loc1}
\begin{tabular}{lrrr}
\toprule
 Loss Function &  MAE &  GAME  \\
\midrule
    Point Loss & 5.97 &  7.24  \\
    LCFCN Loss & 1.15 &  2.09  \\
 CB LCFCN (Ours) Loss & \textbf{0.66} &  \textbf{1.74}  \\
\bottomrule
\end{tabular}
\end{table}

\begin{table}
\centering
\caption{COVID-19-C-Mixed Counting and Localization}
\label{tab:loc2}
\begin{tabular}{lrrr}
\toprule
 Loss Function &  MAE &  GAME  \\
\midrule
    Point Loss & 9.63 & 11.76  \\
    LCFCN Loss & 1.01 &  1.70  \\
 CB LCFCN Loss (Ours) & \textbf{0.82} &  \textbf{1.42}  \\
\bottomrule
\end{tabular}
\end{table}

\section{Conclusion}
\label{sec:conclusion}
Machine learning has the potential to solve a number challenges associated with COVID-19. One example is the identification of high-risk patients by segmenting infected regions in CT scans. However, conventional annotations methods rely on per-pixel labels which are costly to collect for CT scans. In this work, we have proposed an efficient method that can learn from point-level annotations, which are much cheaper to acquire than per-pixel labels. Our method uses a consistency-based loss that significantly improves the segmentation performance compared to conventional point-level loss on 3 COVID-19 open-source datasets. Further, our method obtained results that almost match the performance of the fully supervised methods and they are more robust against different splits of the data.

{\small
\bibliographystyle{abbrvnat}
\bibliography{arxiv}
}

\end{document}

%% file: math_commands.tex
\usepackage{amsmath,amsfonts,bm}

\def\eqref#1{equation~\ref{#1}}

\def\1{\bm{1}}

\DeclareMathAlphabet{\mathsfit}{\encodingdefault}{\sfdefault}{m}{sl}
\SetMathAlphabet{\mathsfit}{bold}{\encodingdefault}{\sfdefault}{bx}{n}

